# The generalisation of the DMCA coefficient to serve distinguishing between hedge and safe haven capabilities of the gold

Mohamed Arbi Madani[a] and Zied Ftiti[b]

[a]*University of Tunis, ISG-T, LR GEF-2A, Tunis, Tunisia;*

[b]*EDC Paris Business School, OCRE-Laboratory, Paris, France.*

## Abstract

This study aims to investigate the role of gold as a hedge and/or safe haven against oil price and currency market movements for medium (calm period) and large (extreme movement) fluctuations. In revisiting the role of gold, our study yields new insights in the literature. First, our empirical design relaxes the assumption of homogeneous investors in favour of agents with different horizons. Second, we develop a new measure of correlation based on the multifractal approach, called the q-detrending moving average cross-correlation coefficient. This allows us to measure the dependence for calm and extreme movements. The proposed measure is both time varying and time-scale varying, taking into account the complex pattern of commodities and financial time series (chaotic, non-stationary, etc.). Using intraday data from 23 May 2017 to 12 March 2019, including 35608 observations for each variable, our results are as follows. First, we reveal clearly that gold acts as a weak (and strong) hedge for oil (and currency) market movements across all type of agents. Second, the safe-haven properties of gold are further established for different horizons. Then, we examine the hedging and stabilising benefits of gold over calm and turmoil periods for gold–oil futures and gold–currency portfolios by estimation of the optimal portfolio weights and the optimal hedge ratio. We confirm the usefulness of gold for hedging and safe havens at different investment horizons, which favours the inclusion of gold futures in oil futures and currency portfolios for risk management purposes.

# 1. Introduction

Quantifying and developing "null-hypothesis" tests in order to check the statistical validity of correlation matrices are two essential steps for understanding the complexity of economy and underlying interactions, but also for asset allocation and portfolio risk estimation of financial markets practitioners (Markowitz, 1959; Laloux et al., 1999; Plerou et al., 2002). The pioneers in developing theoretical concept of connection between time series come back to Galton (1885). Then, Pearson (1895) defined the (Galton) correlation coefficient aiming to measure the similarity of price changes between pairs of assets, known since as the Pearson correlation coefficient[1]. Several measures have been developed in order to measure the cross-correlation between time series and especially since 80s, period of econometric theory revolution. Most of them deal with linear models and stationary time-series (Engle and Granger, 1987; Engle and Granger; 1991; etc.). Some other studies deal with time-varying connection and extended correlation matrix in different ways in order to extract useful information to understand time-varying financial markets dependence, (Bollerslev et al, 1988; Ding and Engle, 2001; Campbell et al. 2008, Forbes and Rigobon 2002; Krishan et al. 2009, Huang et al. 2013). Since the last decade, a new strand of literature highlighting more complexity behavior of financial series, especially after the occurrence of various turmoil periods in short time frame (since 2000). Interestingly, these studies point out the high degree of the non-stationary behavior of financial series, the self-affinity behavior, which might characterize cross-correlation by power-laws.

Based on the previous studies, several techniques have been developed to study fractal and multifractal properties in time series. Among them, the detrended fluctuation analysis noted (DFA) of Peng et al., (1995), and the detrending moving average analysis noted (DMA) of Vandewalee and Ausloos(1998) and their multifractal version, the multifractal detrended fluctuation analysis (MFDFA) of Kantelhardt et al. 2002 and the multifractal detrending moving average (MFDMA) of Gu and Zhou, 2010, respectively. Accordingly, these methods can only be used to study auto-correlation of a single time series, instead of cross correlation between two time series. Until Podobnik et al., 2008, successfully proposed a new method called the detrended cross-correlation analysis (DCCA) related to fractal theory to investigate power-law cross correlations between different simultaneously recorded time series in the presence of nonstationarity. Podobnik's work brought a lot of interest among academicians and practitioners[2], and as a next step a scale-specific correlation coefficient based on DCCA and DMCA have been proposed (Zebende, 2011 and Kristoufek, 2014b). The DMCA method is different in two important aspects; i) it is not based on the box-splitting procedure and ii) assumes a power-law scaling of covariances with an increasing moving average window size s. In other words, The DMCA method can be seen as an improvement of the DCCA approach, but it has the advantage of not requiring dividing the series into nonoverlapping boxes. The DMCA method proposes rather to detrend the series through the subtraction of a continuous function of the series using the moving average.

---

[1] Pearson correlation coefficient has been interpreted in different ways in order to analysis the connection between time series. For more details, please see Rodgers and Nicewanders (1988).

The two methods discussed above conduct to deal with a pair of nonstationary signals, which can be produced both by the long-range cross-correlations and by the pdf's heavy tails. They are seen as an analogue of the Pearson coefficient since they are calculated in the same way as the variance and the covariance analysis[3]. The main limit of these methods is that they can only be used to quantify the cross correlations between any signals and that for only the second moment. Allowing for instability and chaotic structural changes in commodities and financial markets, it is important to develop more sophisticated quantitative measures of the dependency structure for any amplitude and specially for high fluctuations (e.g. third and fourth moments).

In order to avoid such insensitivity, this paper proposes a generalization of the detrending cross-correlation moving average coefficient and it is used here to investigate the capacity of gold as hedge and safe haven against oil and exchange rate from a multi-scale horizon. Comparing with the related literature, we present the contributions of this paper below.

First, previous empirical research has examined for one or, at the most, two time scales (the short and the long run) and using different econometric techniques (e.g., cointegration theory or the vector autoregressive model), little is known about how gold-oil and gold-exchange rates co-move in different time scales. This paper fills this gap by evaluating the ability of gold to hedge and safe haven against oil and USD depreciation at different time scales by using the q-detrending moving average cross-correlation coefficient (q-DMCA) developed in the next section. Second, the proposed q-DMCA coefficient is compared to the q-DCCA coefficient developed by Kwapien et al. (2015). Based on detrending moving average cross-correlation values at different time scales, we propose a bootstrap test to identify the ability of gold to hedge and safe haven against oil and exchange rates that consists of non-overlapping confidence intervals for the level of correlation coefficients; medium and high fluctuations. Finally, the estimate results are afterwards used to test the significant role of gold as a hedge and safe haven against oil and USD depreciation at different time scales by estimating the optimal portfolio weights and the optimal hedge ratio. Using these results, we explore the hedging and stabilizing asset of gold over calm and turmoil periods for a gold-oil futures and gold-currency portfolios.

---

[2] For more details about the applications of the detrended cross-correlation analysis (DCCA) see for example, Podobnik et al., 2009; Zebende and Filho, 2009; Siqueira et al., 2010; Wang et al., 2013; Ruan et al., 2016.

[3] In an analogous way, the variance is presented by the detrended variance function ( $F^2_{DFA}$ or $F^2_{DMA}$ ) and the covariance is presented by the detrended covariance function ( $F^2_{DCCA}$ or $F^2_{DMCA}$ ).

The rest of the paper is laid out as follows: Section 2 presents a brief literature review. In Section 3, we develop the q-DMCA coefficient that depends on the exponent q and the temporal scale s. In section 4, we validate the q-DMCA coefficient and it was compared with the q-DCCA coefficient by using numerical experiments with the mixed-correlated ARFIMA processes. Section 5 presents the main empirical results for the ability of gold to hedge and safe haven against exchange rate and oil price and the intraday attractiveness of gold by using optimal portfolio weights and hedge ratio. Section 6 concludes the paper.

## 2. Related literature review

Several studies have examined the potential role of gold as a hedge or investment safe haven against inflation, stock markets, oil and USD depreciation[4]. Initially, to examine the relationship between gold and USD exchange rate, researchers used linear cross-correlation measures. By examining gold's holding positions for US and non-US investors, Beckers and Soenen (1984) found a negative correlation between the return on gold investments (expressed in US dollars) and the strength of the US dollar on the foreign exchange market and an asymmetric risk diversification with advantage for non-US investors. Sjasstad and Scacciavillani (1996) and Sjasstad (2008) concluded that the appreciation or depreciation of the dollar has significant effects on gold price by using the forecast error approach. Capie et al. (2005) have used GARCH models to confirm the hedging power of gold against the USD. Using DCCA-GARCH model, Joy (2011) indicates that gold is a weak safe haven and a successful hedge against the USD. Furthermore, the dependence structure of gold with exchange rates and stock returns has been examined by considering the marginal impact of stock returns on gold returns using a threshold regression model, with the threshold given by a specific quantile of the stock return distribution (see, e.g., Baur and Lucey, 2010; Baur and McDermott, 2010; Wang and Lee, 2011; Ciner et al., 2013). More recently, Chang et al. (2013) investigated the correlation among oil prices, gold prices and NT dollar versus U.S. dollar exchange rate by employing several linear tests and models (Johansen co-integration test, Granger causality test, VAR model, impulse response analysis and variance decomposition method) and conclude that the variables are considerably independent. Reboredo and Rivera-Castro (2014b) used the peaks-over-threshold (POT) approach to identify extreme gold and USD exchange rate values and concluded that gold can serve as a hedge against US dollar depreciation but is a weak safe haven against extreme US dollar movements.

---

[4] Other studies have examined the role of gold as a hedging device against inflation (Chua and Woodfward, 1982; Tully and Lucey, 2007; Blose, 2010; Wang et al, 2011; Van Hoang et al, 2016; Lucey et al, 2017; Beckmann et al, 2019) and stock market movements (Baur and Lucey, 2010; Baur and McDermott, 2010; Miyazaki et al., 2012; Ftiti et al, 2016; Nguyen et al, 2016).

Moreover, another array of research has been reviewed and findings suggest strong relationships between gold and oil prices (Cashin et al., 1999; Ye, 2007; Zhang et al., 2007). Using GARCH family models, Hammoudeh and Yuan (2008) examined the volatility behavior of three metals: gold, silver and copper and found that oil shock does not impact all three metals similarly. Focusing on oil and six metal prices, Lescaroux (2009) found that the tendency of commodity prices to oscillate together reflects the tendency of their fundamental factors to move together. Soytas et al. (2009) investigated the long- and short-run transmissions of information between the world oil price, domestic spot gold and silver price. They concluded that no causal relationship amongst the variables. Other studies have suggested a long-term relationship between oil and gold prices (Narayan et al., 2010; Zhang and Wei, 2010; Šimáková, 2011; Jain and Ghosh, 2013; Bouri et al. 2017).

The aforementioned studies examine the co-movement between gold, oil and exchange rate by static way; in other words, their empirical analyses are based on the assumption of stability in long run relationships. However, since there is a common phenomenon that structural breaks often exist in economic and financial markets. Therefore, this assumption is not reasonable. Far away from the normal cycle theory, financial and commodity markets are characterized by chaotic structural change since the stock market crash of October 19, 1987 (Hsieh, 1991) and can be assisted by evolutionary and complex systems approaches, especially ones that privilege the role of interactive knowledge and belief systems. As a matter of fact, financial globalization since the 1970s and dynamic patterns in the global economy are caused by developments in information-processing technologies; government deregulation; and the more global nature of all economic activity. There was a continuing rapid expansion of international financial activity, continuing at least to a peak in 2006 of 'the long boom' (The Joseph effect) that preceded the recent global crisis as late as December 2008 (The Noé effect), generating $6 trillion or approximately 20 percent of world GDP.

In order to specify these more general mean structures in gold price, oil price and exchange rate relationships, many authors have employed nonlinear models. By applying a structural break cointegration test of Gregory and Hansen (1996), Narayan et al. (2010) confirmed a structural break cointegration between the mentioned markets. Reboredo (2013) used copulas to characterize average and extreme market dependence between gold and the USD and empirical results suggest that gold can act as hedge and safe haven against dollar depreciation. Kanjilal and Ghosh (2017) employed the threshold cointegration to find a non-linear relationship between gold and oil prices. The nonlinear ARDL model has been also employed by Bildirici and Turkmen (2015) and kumar (2017) to underline the importance of asymmetric co-movement between gold and oil markets. Reboredo and Rivera-Castro (2014b) and Baruník et al. (2016) examined the gold-exchange rate and gold-oil relationships respectively from a different-investment-horizons perspective by using wavelet approach.

## 3. The generalisation of the DMCA: The q-DMCA coefficient

We can use the information provided by the detrending cross-correlation moving average analysis to distinguish between hedge and safe-haven properties which measure dependence between two or more variables in terms of average movements by the second order ($q=2$) and in terms of extreme market movements by the fourth order ($q=4$). According to the definitional approach described in Kaul and Sapp (2006), Baur and Lucey (2010) and Baur and McDermott (2010), the distinctive features of an asset as a hedge or safe haven are as follows:

• Hedge: an asset is a weak (or strong) hedge if it is uncorrelated (or negatively) correlated with another asset or portfolio on average.

• Safe haven: an asset is a weak (or strong) safe haven if it is uncorrelated (or negatively) correlated with another asset or portfolio in times of extreme market movements.

The DMCA coefficient can be seen as an alternative and a complement to the DCCA coefficient (Kristoufek, 2014b). According to results found by Kristoufek (2014a) and Sun and Liu (2016), the DCCA coefficient dominates the Pearson's coefficient for possibly non-stationary series and the DMCA coefficient which, in turn, dominates the already mentioned coefficients (DCCA and Pearson's coefficient). This new measure is based on the detrending moving average (DMA) (Vandewalle and Ausloos (1998); Alessio et al. 2002).

For two possibly non-stationary series $\{x_t\}$ and $\{y_t\}$, we construct the cumulative sum $X_t = \sum_{i=1}^{t} x_i$ and $Y_t = \sum_{i=1}^{t} y_i$ for t = 1, 2, ..., N. Where N is the length for both series (the two time series have the same length, N). According to Alessio et al. (2002), the moving average functions $\widetilde{X}_t$ and $\widetilde{Y}_t$ are defined as

$$\widetilde{X}_t = \frac{1}{n} \sum_{k=-\lfloor (n-1)\theta \rfloor}^{\lceil (n-1)(1-\theta) \rceil} X(t-k), \tag{1}$$

$$\widetilde{Y}_t = \frac{1}{n} \sum_{k=-\lfloor (n-1)\theta \rfloor}^{\lceil (n-1)(1-\theta) \rceil} Y(t-k), \tag{2}$$

where the position parameter θ varies from 0 to 1. The reference point θ of the moving average is set in the sliding window n. $\lfloor x \rfloor$ is the largest integer not greater than x, and $\lceil x \rceil$ is the smallest integer not less than x. Varying θ, the moving average function $\widetilde{y}(t)$ contains different information. Three cases are possible for setting the parameter θ[5] (Arianos et Carbone, 2007): i) θ = 0 refers to the backward moving average which depends only on the past points of the time series, ii) θ = 0.5 refers to the centered moving average which is obtained by half-past and half-future data points, and iii) θ = 1 refers to the forward moving average which depends only on the future data points. The residual series is obtained by subtracting the trend $\widetilde{X}(i)$ from $X(i)$, $\varepsilon_X(i) = X(i) - \widetilde{X}(i)$ and in the same way we have $\varepsilon_Y(i) = Y(i) - \widetilde{Y}(i)$ where $n - \lfloor (n-1)\theta \rfloor \leq i \leq N - \lfloor (n-1)\theta \rfloor$. By dividing the residual series into Ns parts of equal size s, where Ns corresponds to the integer part of (N/s−1). Define each part as $\epsilon_v$, so that $\epsilon_v(i) = \epsilon(l + i)$, for $1 \leq i \leq n$, where l = (v - 1)s. We can calculate the root-mean-square function fv(s) with the segment size s by

$$f_v^2 = \frac{1}{s} \sum_{i=1}^{n} \varepsilon_v^2(i). \tag{3}$$

Finally, the overall detrended fluctuation functions of each time series are estimated as follows

---

[5] In this study, we follow Shao *et al.* (2012), who used the centred moving average (θ = 0.5), as this leads to the best solution.

$$F_{X,DMA}^2 = \frac{1}{N_s} \sum_{v=1}^{N_s} f_{X,v}^2(s), \tag{4}$$

$$F_{Y,DMA}^2 = \frac{1}{N_s} \sum_{v=1}^{N_s} f_{Y,v}^2(s), \tag{5}$$

The bivariate fluctuation function $F_{DMCA}^2$ is defined in line to Kristoufek (2014),

$$F_{DMCA}^2(s) = \frac{1}{s} \sum_{k=-\lfloor (s-1)\theta \rfloor}^{\lceil (s-1)(1-\theta) \rceil} (X_t - \tilde{X}_t)(Y_t - \tilde{Y}_t) \tag{6}$$

The DMCA coefficient can be easily obtained by following Zebende (2011) for the DCCA coefficient,

$$\rho_{DMCA}(s) = \frac{F_{DMCA}^2(s)}{F_{X,DMA}(s) F_{Y,DMA}(s)}. \tag{7}$$

According to the Cauchy–Schwarz inequality, we have $-1 < \rho_{DMCA}(s) < 1$. When $\rho_{DMCA} = 0$, no cross-correlation between the two time series $\rho_{DMCA} = -1$ means that the bivariate time series possesses perfect long-range negative cross-correlation and $\rho_{DMCA} = 1$ means that the bivariate time series possesses perfect long-range cross-correlation.

As we can see in the above equations, the DMCA coefficient is represent as the ratio between the detrended covariance function $F_{DMCA}^2$ and the detrended variance function $F_{DMA}^2$. This fact takes into account only the level of cross-correlation in mean and makes the measure unsuitable for the other amplitudes. In other words, the values of $\rho_{DMCA}$ may not be the same for all fluctuations (lower q<0 and higher q>0).

In order to surpass this limit, we propose a multifractal generalization of the detrending moving average cross-correlation coefficient. The idea is the same as that of Kwapien et al. (2015), which consists in making the coefficient DMCA to the power q, so that it becomes more attractive by made it to depend on the exponent q and the temporal scale s. Our new measure is based on the so-called the q[th] order fluctuation function Fq from the MFDMA and MF-X-DMA methods (Gu and Zhou, 2010; Jiang and Zhou, 2011). Therefore, we use the detrended covariance sign, which allows to keep "all" information about analyzed time series (Oswiecimka et al. 2014). These quantities are defined respectively as follows:

$$F^q(s) = \frac{1}{N_s} \sum_{v=1}^{Ns} f_v^q(s), \tag{8}$$

$$F^q{}_{xy}(s) = \frac{1}{N_s} \sum_{\upsilon=1}^{N_s} sign[F_\upsilon(s)] |F_\upsilon(s)|^{\frac{q}{2}}, \tag{9}$$

where, $F_\upsilon(s) = \frac{1}{s} \sum_{k=1}^{s} [X_\upsilon(k) - \widetilde{X}_\upsilon(k)][Y_\upsilon(k) - \widetilde{Y}_\upsilon(k)]$. (10)

Finally, we propose the new detrending moving average $q^{th}$ order cross-correlation coefficient (q-DMCA cross-correlation coefficient) as follows,

$$\rho_{q-DMCA}(s) = \frac{F^q_{xy}(s)}{\sqrt{F^q{}_x(s) F^q{}_y(s)}}. \tag{11}$$

For q > 0, according to the Cauchy-Schwarz inequality we have,

$$-1 < \rho_{q-DMCA} < 1. \tag{12}$$

When q < 0, the coefficient q-DMCA in absolute value takes values greater than 1, frequently when the bivariate series are not cross-correlated or weakly cross-correlated. In order to take this case into consideration, the q-DMCA coefficient can be redefined as follows:

$$\rho^*{}_{q-DMCA}(s) = \begin{cases} \rho_{q-DMCA}(s) \text{ if } |\rho_{q-DMCA}(s)| \leq 1 \\ [\rho_{q-DMCA}(s)]^{-1} \text{ if } |\rho_{q-DMCA}(s)| > 1. \end{cases} \tag{13}$$

## 4. Numerical experiments with the mixed-correlated ARFIMA processes

In order to evaluate our q-DMCA coefficient, we compare it against the q-DCCA coefficient by using a numerical experiment based on mixed-correlated ARFIMA (MC-ARFIMA) processes which allows for various specifications of univariate and bivariate long-term memory (Kristoufek, 2013). The two methods focus to estimate the generalized power-law coherency parameter $H_\rho(q)$ via controlling the generalized univariate Hurst exponents $H_x(q)$, $H_y(q)$ and the generalized bivariate Hurst exponent $H_{xy}(q)$. The parameter $H_\rho$ is defined as $H_\rho(q) = H_{xy}(q) - 1/2(H_x(q) + H_y(q))$.

The MC-ARFIMA processes are defined as:

$$x_t = \alpha \sum_{n=0}^{+\infty} y_i a_n(d_1) \varepsilon_{1,t-n} + \beta \sum_{n=0}^{+\infty} a_n(d_2) \varepsilon_{2,t-n} \tag{14}$$

$$y_t = \gamma \sum_{n=0}^{+\infty} y_i a_n(d_3) \varepsilon_{3,t-n} + \delta \sum_{n=0}^{+\infty} a_n(d_4) \varepsilon_{4,t-n} \tag{15}$$

For specific $d_i = H_i - 0.5$ we define $a_n(d_i)$ as

$$a_n(d) = \frac{\Gamma(n+d)}{\Gamma(n+1)\Gamma(d)} \tag{16}$$

The innovations are characterized by

$$\langle \varepsilon_{i,t} \rangle = 0 \text{ for } i = 1,2,3,4$$

$$\langle \varepsilon^2_{i,t} \rangle = \sigma^2_{\varepsilon_i} \text{ for } i = 1,2,3,4$$

$$\langle \varepsilon_{i,t} \varepsilon_{j,t-n} \rangle = 0 \text{ for } n \neq 0 \text{ and } i,j = 1,2,3,4$$

$$\langle \varepsilon_{i,t} \varepsilon_{j,t} \rangle = \sigma_{ij} \text{ for } i \neq j \text{ and } i,j = 1,2,3,4. \tag{17}$$

Specially, we have $H_x = d_1 + 0.5$, $H_y = d_4 + 0.5$ and $H_{xy} = 0.5 + 1/2(d_2 + d_3)$. In the simulations, we initialize the following parameters: $d_1 = d_4 = 0.4$, $d_2 = d_3 = 0.2$ and the theoretical values of Hust exponents and power coherency parameter are thus equal to $H_X = H_y = 0.9$, $H_{xy} = 0.7$ and $H_\rho = -0.2$. Following Kristoufek (2017), we set three different simulated time series lengths $T = 500, 1000, 5000$. We remind that the power-law coherency is obtained when $\{\varepsilon_2\}$ and $\{\varepsilon_3\}$ are correlated, for that we study three correlation levels 0.1, 0.5 and 0.9. For each correlation level we simulate 1000 bivariate series[6]. In order to obtain the estimated values of $H_\rho(q)$ we use the variance and covariance scales relations:

$$F_x(q,s) \sim s^{H_x(q)}, \tag{18}$$

$$F_y(q,s) \sim s^{H_y(q)}, \tag{19}$$

$$F_{xy}(q,s) \sim s^{H_{xy}(q)}, \tag{20}$$

Then we obtain the generalized scaling squared correlation as follow

$$\rho^2_{q-DMCA}(s) \sim \frac{s^{2qH_{xy}(q)}}{s^{qH_x(q)} s^{qH_y(q)}} = s^{2qH_{xy}(q) - qH_x(q) - qH_y(q)} = s^{2qH_\rho(q)}. \tag{21}$$

The estimated values of $H_\rho(q)$ are easy obtained by using the log-log regression. Simulation results are performed according to three criteria: bias, variance and mean squared error (MSE, the sum of squared bias and variance) of the estimators.

In our work we are only interested on the order $q = 2$ and $q = 4$, as explained previously in section 3.

---

[6] We have set the same setting in Réf. Kristoufek(2017), in order to assess the introduction of the sign function in the detrended covariance function.

**Table 1.** Simulation results of DCCA Method (for q=2)

| q=2 | | ρ=0,1 | | | ρ=0,5 | | | ρ=0,9 | | |
|---|---|---|---|---|---|---|---|---|---|---|
| | | bias | SD | MSE | bias | SD | MSE | bias | SD | MSE |
| N=500 | $n_{min}$=10 | 0.4611 | 0.5188 | 0.4818 | 0.1099 | 0.1271 | 0.0282 | 0.0319 | 0.0376 | 0.0024 |
| | $n_{min}$=20 | 0.2798 | 0.3355 | 0.1908 | 0.0893 | 0.1411 | 0.0279 | 0.0499 | 0.0577 | 0.0058 |
| | $n_{min}$=50 | 0.1271 | 0.1835 | 0.0498 | 0.0915 | 0.1376 | 0.0273 | 0.0726 | 0.0907 | 0.0135 |
| N=1000 | $n_{min}$=10 | 0.4984 | 0.5512 | 0.5522 | 0.104 | 0.1125 | 0.0235 | 0.0304 | 0.0347 | 0.0021 |
| | $n_{min}$=50 | 0.1534 | 0.1976 | 0.0626 | 0.0751 | 0.1206 | 0.0202 | 0.0608 | 0.071 | 0.0087 |
| | $n_{min}$=100 | 0.0919 | 0.1348 | 0.0266 | 0.0785 | 0.1092 | 0.0181 | 0.0785 | 0.0983 | 0.0158 |
| N=5000 | $n_{min}$=10 | 0.4751 | 0.4964 | 0.4721 | 0.1003 | 0.1021 | 0.0205 | 0.0279 | 0.0293 | 0.0016 |
| | $n_{min}$=50 | 0.2335 | 0.2748 | 0.13 | 0.0332 | 0.0499 | 0.0036 | 0.0552 | 0.058 | 0.0064 |
| | $n_{min}$=100 | 0.1441 | 0.1837 | 0.0545 | 0.0514 | 0.0808 | 0.0092 | 0.0563 | 0.0616 | 0.007 |

**Table 2.** Simulation results of DMCA Method (for q=2)

| q=2 | | ρ=0,1 | | | ρ=0,5 | | | ρ=0,9 | | |
|---|---|---|---|---|---|---|---|---|---|---|
| | | bias | SD | MSE | bias | SD | MSE | bias | SD | MSE |
| N=500 | $s_{max}$=20 | 0.2836 | 0.3362 | 0.1934 | 0.1155 | 0.1666 | 0.0411 | 0.0403 | 0.054 | 0.0045 |
| | $s_{max}$=50 | 0.1232 | 0.1789 | 0.0472 | 0.1014 | 0.1519 | 0.0333 | 0.0735 | 0.0979 | 0.015 |
| | $s_{max}$=100 | 0.2629 | 0.3555 | 0.1955 | 0.0836 | 0.1127 | 0.0197 | 0.0944 | 0.1233 | 0.0241 |
| N=1000 | $s_{max}$=20 | 0.323 | 0.3685 | 0.2401 | 0.0923 | 0.1176 | 0.0223 | 0.0309 | 0.0378 | 0.0024 |
| | $s_{max}$=50 | 0.1522 | 0.2031 | 0.0644 | 0.0879 | 0.133 | 0.0254 | 0.0582 | 0.0794 | 0.0097 |
| | $s_{max}$=100 | 0.0893 | 0,131 | 0.0251 | 0.0824 | 0.1135 | 0.0197 | 0.0818 | 0.1061 | 0.0179 |
| N=5000 | $s_{max}$=20 | 0.3854 | 0.4219 | 0.3265 | 0.0814 | 0.0867 | 0.0141 | 0.0211 | 0.0242 | 0.001 |
| | $s_{max}$=50 | 0.2428 | 0.2822 | 0.1386 | 0.0515 | 0.0691 | 0.0074 | 0.0353 | 0.0398 | 0.0028 |
| | $s_{max}$=100 | 0.1539 | 0.1972 | 0.0626 | 0.0631 | 0.098 | 0.0136 | 0.0427 | 0.0507 | 0.0044 |

The Simulation results for q=2 of the q-DMCA and q-DCCA methods are presented in Table 1 and Table 2 respectively. First, we can deduce that the detrended covariance sign brings a significant improvement of the results comparing with findings of Kristoufek (2017), especially for the q-DMCA analysis.

Second, for low correlation between error-terms $\varepsilon_2$ and $\varepsilon_3$, the bias achieves roughly 0.5 and 0.4 for the DCCA and DMCA based methods respectively. The situation improves much more when the correlation between $\varepsilon_2$ and $\varepsilon_3$ increases, chiefly due to very low variance of the estimators.

The best case is when the correlation between innovations is equal to 0.9 and the window size is the shortest $s_{min}$=10 and $n_{max}$=20 for the estimators q-DCCA and q-DMCA respectively. We can generally say that the bias and variance decrease with time series length, for a sample of 5000 observations we have approximately 0.03 (bias) and 0.03 (SD) for the DCCA method and 0.02 (bias) and 0.02 (SD) for the DMCA method.

The simple change made to the detrended covariance by introducing the sign adduces a mainly advantage in term of bias and variance, this fact can be explained by the ability of estimators to capture almost all information.

We can conclude from Table 1 and 2 that for both methods when the correlation between error-terms $\varepsilon_2$ and $\varepsilon_3$ increases, the results improve significantly.

**Table 3.** Simulation results of DCCA Method (for q=4)

| q=4 | | $\rho=0,1$ | | | $\rho=0,5$ | | | $\rho=0,9$ | | |
|---|---|---|---|---|---|---|---|---|---|---|
| | | bias | SD | MSE | bias | SD | MSE | bias | SD | MSE |
| N=500 | $n_{min}$=10 | 0.0217 | 0.0271 | 0.0012 | 0.0353 | 0.0412 | 0.0029 | 0.0794 | 0.0821 | 0.013 |
| | $n_{min}$=20 | 0.0458 | 0.0526 | 0.0049 | 0.062 | 0.0676 | 0.0084 | 0.0925 | 0.0957 | 0.0177 |
| | $n_{min}$=50 | 0.0808 | 0.0874 | 0.0142 | 0.0888 | 0.0948 | 0.0169 | 0.1041 | 0.1088 | 0.0227 |
| N=1000 | $n_{min}$=10 | 0.0168 | 0.0212 | 0.0007 | 0.0346 | 0.039 | 0.0027 | 0.081 | 0.0823 | 0.0133 |
| | $n_{min}$=50 | 0.0814 | 0.085 | 0.0138 | 0.0911 | 0.0942 | 0.0172 | 0.108 | 0.1103 | 0.0238 |
| | $n_{min}$=100 | 0.099 | 0.1029 | 0.0204 | 0.1056 | 0.1094 | 0.0231 | 0.1154 | 0.1186 | 0.0274 |
| N=5000 | $n_{min}$=10 | 0.0095 | 0.0115 | 0.0002 | 0.0355 | 0.0365 | 0.0026 | 0.0816 | 0.0819 | 0.0134 |
| | $n_{min}$=50 | 0.0851 | 0.0859 | 0.0146 | 0.0945 | 0.0952 | 0.018 | 0.1115 | 0.112 | 0.025 |
| | $n_{min}$=100 | 0.1019 | 0.1029 | 0.021 | 0.1076 | 0.1085 | 0.0233 | 0.1179 | 0.1187 | 0.028 |

**Table 4.** Simulation results of DMCA Method (for q=4)

| q=4 | | $\rho=0,1$ | | | $\rho=0,5$ | | | $\rho=0,9$ | | |
|---|---|---|---|---|---|---|---|---|---|---|
| | | bias | SD | MSE | bias | SD | MSE | bias | SD | MSE |
| N=500 | $s_{max}$=20 | 0.0385 | 0.045 | 0.0035 | 0.0502 | 0.0559 | 0.0056 | 0.0766 | 0.0804 | 0.0123 |
| | $s_{max}$=50 | 0.0727 | 0.0791 | 0.0115 | 0.0786 | 0.0852 | 0.0134 | 0.094 | 0.0992 | 0.0187 |
| | $s_{max}$=100 | 0.0866 | 0.0952 | 0.0166 | 0.0912 | 0.0996 | 0.0182 | 0.1008 | 0.1084 | 0.0219 |
| N=1000 | $s_{max}$=20 | 0.0357 | 0.04 | 0.0029 | 0.05 | 0.0537 | 0.0054 | 0.078 | 0.08 | 0.0125 |
| | $s_{max}$=50 | 0.0733 | 0.0773 | 0.0113 | 0.0816 | 0.0854 | 0.0139 | 0.0965 | 0.0994 | 0.0192 |
| | $s_{max}$=100 | 0.0901 | 0.0948 | 0.0171 | 0.0967 | 0.101 | 0.0195 | 0.105 | 0.1091 | 0.0229 |
| N=5000 | $s_{max}$=20 | 0.0372 | 0.0381 | 0.0028 | 0.0525 | 0.0533 | 0.0056 | 0.0803 | 0.0807 | 0.013 |
| | $s_{max}$=50 | 0.0762 | 0.0769 | 0.0117 | 0.0844 | 0.0852 | 0.0144 | 0.0992 | 0.0998 | 0.0198 |
| | $s_{max}$=100 | 0.0956 | 0.0964 | 0.0184 | 0.1004 | 0.1012 | 0.0203 | 0.1091 | 0.1099 | 0.024 |

In the same way, simulation results for q=4 of the q-DMCA and q-DCCA coefficients are presented in Table 3 and Table 4 respectively. First, we can remark that the two methods give a well estimation of the parameter $H_\rho$ when we focus on the analyses of high fluctuations. Second, findings are independent of the level of correlation and the sample length. In other words, when we change the setting, results remain relevant. Similarly, the best situation for high fluctuations is when the correlation between $\varepsilon_2$ and $\varepsilon_3$ is lower ($\rho=0,1$).

In order to improve our numerical study, we extend the length of samples until 100000 observations. The purpose of this fact is to verify whether the methods are stable with respect to the length of the sample and for more comparison between them (check if the alignment found between the two methods remains valid).

Figure 1 shows the estimation values of the parameter $H_\rho$ against time series length. The main deduction from results is that the two methods show a stable estimation (in mean) regardless the length of sample. For q=2, the estimation values are approximately equal to -0.15 and -0.16 for DCCA and DMCA methods, respectively. For q=4, we have almost the same result as shown by the dashed lines in figure1. Comparing with the theoretical value which is equal to -0.2, we can deduce that the DMCA method is more efficient than its competitor the DCCA method.

**Figure 1.** Comparative analyses between DCCA and DMCA methods

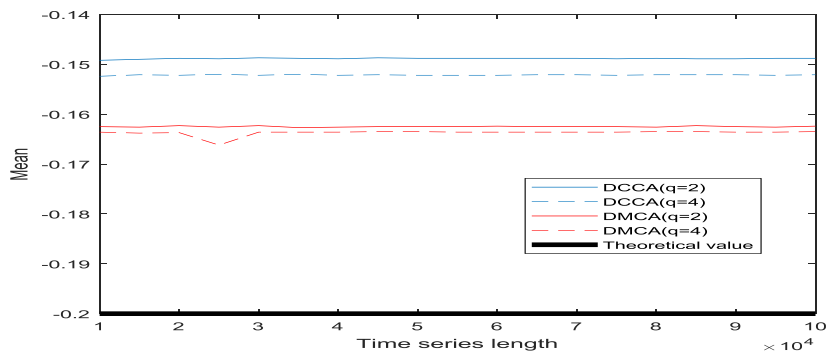

## 5. Empirical validation

### *5.1. Data and preliminary analysis*

This section is devoted to an empirical application of the q-DMCA coefficient to determine the ability of gold to hedge and safe-haven the dollar depreciation. To do that, we use high frequency data (five-minute intervals), which enables us to investigate more important and interesting information and capture further phenomena at short-term intervals. The data related to the gold (expressed in USD per ounce), the light sweet crude oil (expressed in USD per barrel) futures contracts and exchange rates (measured as unit of foreign currency per one USD, where an exchange rate decrease means USD depreciation) were collected from Bloomberg database on an intraday basis and cover a period of approximately 400 trading days beginning on May 23, 2017 and ending on March 12, 2019. Thus, the sample includes 35608 observations for each variable. We analyze the most important currencies: Euro (EUR), pound sterling (GBP), Swiss franc (CHF), Japanese yen (JPY), Canadian dollar (CAD) and Australian dollar (AUD). Figure 2 portrays the gold-oil price and gold price-exchange rate dynamics for the different currencies inspected in our study.

## Figure 2. The dynamic of gold, oil prices and exchange rates

*Fig. 2a. The dynamic of gold and oil prices*

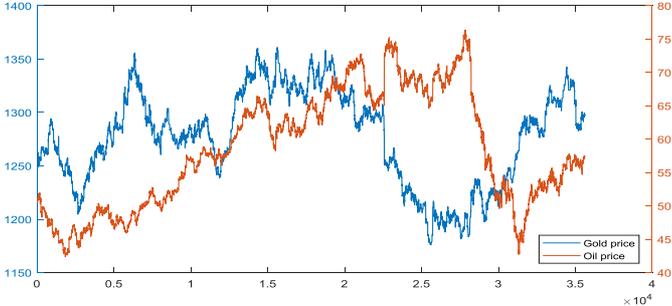

*Fig. 2b. The dynamic of gold price and USD/EUR*

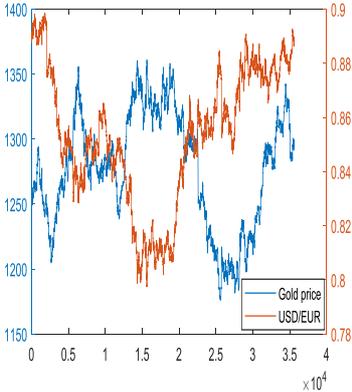

*Fig. 2c. The dynamic of gold price and USD/GBP*

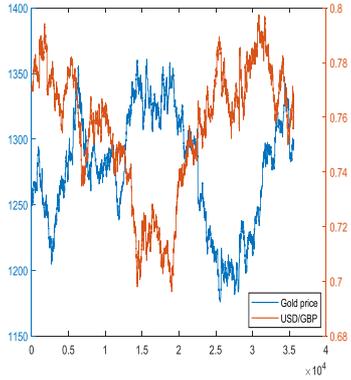

*Fig. 2d. The dynamic of gold price and USD/CHF currency*

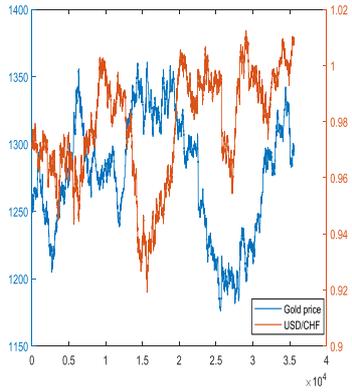

*Fig. 2e. The dynamic of gold price and USD/JPY currency*

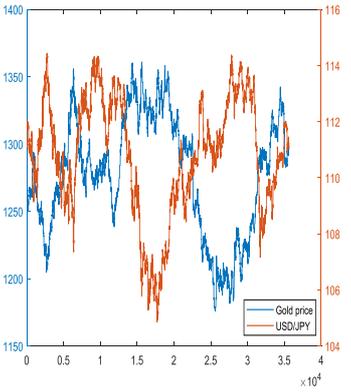

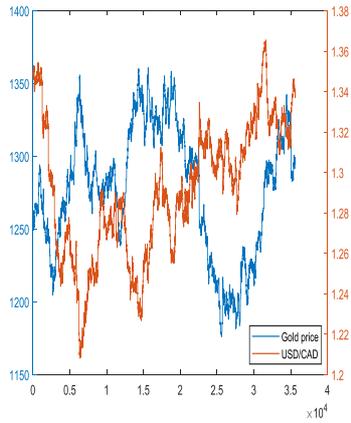
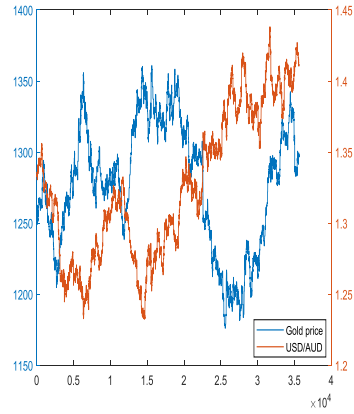

*Fig. 2f. The dynamic of gold price and USD/CAD currency*   *Fig. 2g. The dynamic of gold price and USD/AUD currency*

Table 5 presents the main descriptive statistics for gold, oil and currency return series computed on a continuous compounding basis as the first difference of log prices. The symmetry and normality hypotheses are rejected, suggesting evidence of asymmetry, nonnormality, and leptokurtic excess in the data. This excess might due to extreme movements in the time series under consideration, although figure 2 show repetitive areas of high volatility groupings, suggesting that market turbulence occurs in these series[7]. Further, the rejection of normality suggests the ambiguity to use linear modelling to investigate the hedging and safe haven of gold against oil and dollar depreciation.

**Table 5.** Descriptive statistics of the returns

|  | Gold | Oil | USD/EUR | USD/GBP | USD/CHF | USD/JPY | USD/CAD | USD/AUD |
|---|---|---|---|---|---|---|---|---|
| **Min** | -0.0207 | -0.0703 | -0.0175 | -0.0340 | -0.0136 | -0.0120 | -0.0113 | -0.0168 |
| **Max** | 0.0221 | 0.1104 | 0.0221 | 0.0193 | 0.0165 | 0.0146 | 0.0164 | 0.0148 |
| **Mean** | $7.6653 \cdot 10^{-7}$ | $3.4492 \cdot 10^{-6}$ | $3.7411 \cdot 10^{-8}$ | $2.6678 \cdot 10^{-7}$ | $-1.0117 \cdot 10^{-6}$ | $-1.5614 \cdot 10^{-8}$ | $2.5296 \cdot 10^{-7}$ | $-1.5654 \cdot 10^{-6}$ |
| **S.D** | $7.0407 \cdot 10^{-4}$ | 0.0021 | $5.1261 \cdot 10^{-4}$ | $6.1354 \cdot 10^{-4}$ | $4.6926 \cdot 10^{-4}$ | $4.7494 \cdot 10^{-4}$ | $5.1769 \cdot 10^{-4}$ | $5.8923 \cdot 10^{-4}$ |
| **Skewness** | 0.9573 | 4.4286 | 3.5386 | -4.1958 | 2.0609 | 1.4065 | 2.5621 | -0.8246 |
| **Kurtosis** | 159.9064 | 383.0784 | 252.6336 | 395.7880 | 156.1363 | 143.3555 | 142.1467 | 166.5238 |
| **J.B.(p-value)** | 0.001 | 0.001 | 0.001 | 0.001 | 0.001 | 0.001 | 0.001 | 0.001 |

Note: S.D. and J.B. denote the standard deviation and Jarque–Bera p-value, respectively.

---

[7] The ADF and PP unit root tests reject the stationarity hypothesis for all series. The results are available upon request.

**5.2. does gold is hedge or safe haven**

We applied an extended of the detrending moving average analysis to gold, oil and exchange rate return series using the q-DMCA cross-correlation coefficient for different window lengths from 20 (one hour and half) to 3162 (approximately two month) observations. For each pair of composite variables, we estimated the coefficient $\rho_{q-DMCA}(s)$ for the medium (q=2) and high fluctuations (q=4).

Correlations between gold, oil and exchange rates computed from Eq. (11) are shown in Figure 3. We can note that the cross-correlations between bivariate series are not the same for different window sizes. In other words, the dependences which exist between gold, oil and exchange rate are dependent of the frequency. The main advantage of our analysis is to consider the most of market participants; traders, hedges funds and policymakers, having various horizons.

Regarding the relationship between gold and oil, empirical results indicate independence between the two markets for medium fluctuations (q=2), whatever the time scale. However, they are negatively correlated for high fluctuations (q=4), accepting the safe-haven propriety of the gold during period of turmoil. More specifically for short time scale ($s < 600$), under extreme movement, the dependence between oil and gold is around -15% and close to zero in average for ($s > 600$). This finding shows that gold might be a weak hedge and strong safe haven against the oil price for short time scale.

Concerning the dynamic of the relationship between gold and exchange rates, we observe that correlation is largely more pronounced in the medium fluctuations than large ones. Figs 2b-2g show that the correlation of medium fluctuations for all currency markets is around -35% in average and usually constant across time scale (s). Although that the correlations for large fluctuations (q=4) have a different pattern, the average -20% for all currency markets. These results highlight the capability of gold to hedge as well as to safe haven against exchange rate movements.

**Figure 3.** Dynamic cross-correlation between gold, oil and exchange rate return series

*Figure 3a. Dynamic cross-correlation between gold and oil*

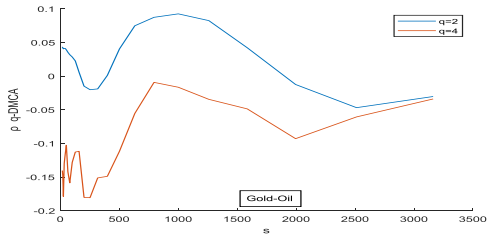

*Figure 3b. Dynamic cross-correlation between gold and EUR exchange rate*

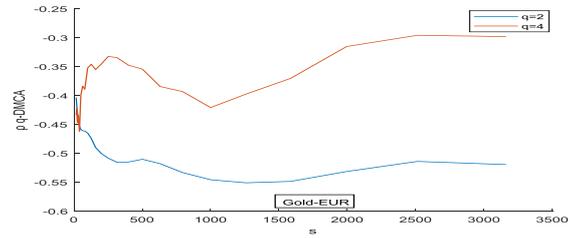

*Figure 3c. Dynamic cross-correlation between gold and GBP exchange rate*

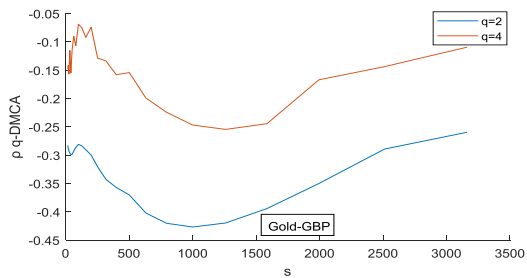

*Figure 3d. Dynamic cross-correlation between gold and JPY exchange rate*

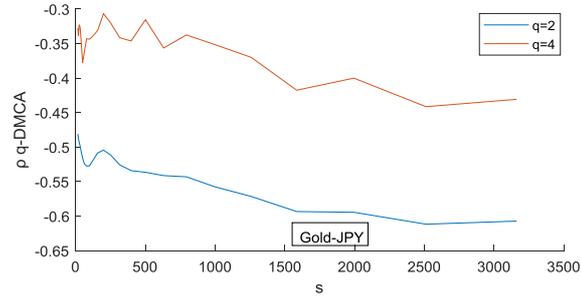

*Figure 3e. Dynamic cross-correlation between gold and CHF exchange rate*

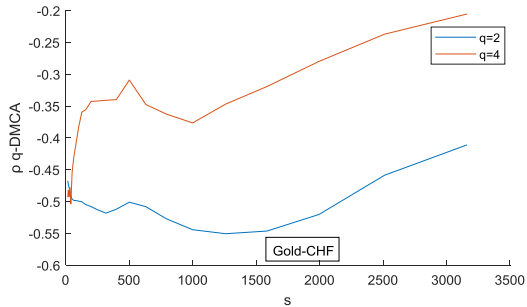

*Figure 3f. Dynamic cross-correlation between gold and CAD exchange rate*

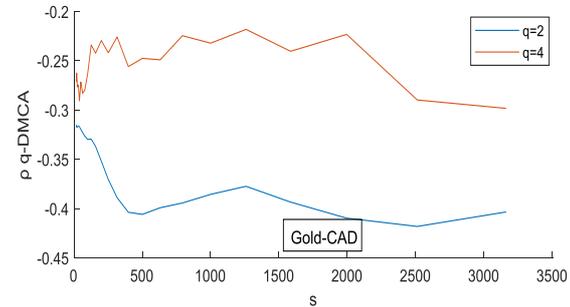

*Figure 3g. Dynamic cross-correlation between gold and AUD exchange rate*

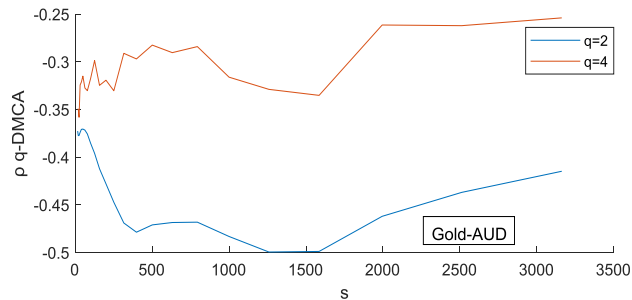

In order to get reliable results, we employ a bootstrap test based on the iterative or improved amplitude adjusted Fourier transform (IAAFT) introduced by Schreiber and Schmitz (1996), which is a well-known method for surrogate data. It is a numerical technique of testing nonlinearity in time series data. The IAAFT algorithm provides a Gaussian series from the original data by the Fourier transform and retains the autocorrelation function sufficiently similar to the original data. This implies the destruction of any correlation that exists between the variables. The implement of the test is easy; the statistic is defined as the q-DMCA coefficient, the distribution of the statistic is generate by an ensemble of the statistic $\rho_{q-DMCA}$, which is obtained by applied the surrogated procedure 1000 times and the q-DMCA coefficient $\rho^s_{q-DMCA}$ of each couple of surrogated series is calculated. The null hypothesis is that the cross-correlation between original series possesses the same dependence traits as those obtained from surrogated series, i.e., $\rho_{q-DMCA} = \overline{\rho}^s_{q-DMCA}$, where $\overline{\rho}^s_{q-DMCA}$ is the mean of all $\rho^s_{q-DMCA}$ values. The difference in term of correlation between the original series and the surrogated series is quantitatively described by a two tailed p-value, which is defined as

$$p = \Pr ob\left(\left|\rho^s_{q-DMCA}(s) - \overline{\rho}^s_{q-DMCA}(s)\right| > \left|\rho_{q-DMCA}(s) - \overline{\rho}^s_{q-DMCA}(s)\right|\right). \qquad (22)$$

If the null hypothesis cannot de rejected, this implies that gold's a weak hedge (safe haven) financial instrument.

The results of the bootstrap test are presented in Tables 6 and 7 for the medium and high fluctuations, respectively. On the one hand, we deduce that gold can act as a significant hedge and safe haven against USD depreciation with differences at different time scales. For the gold–oil relationship, we cannot reject the null hypothesis for average dependence, which leads us to conclude that the two markets are independent in calm periods. On the other hand, there is negative and significant tail dependence between gold and oil for short time scales (less than 600). Our results reveal that gold can hedge against oil price movements weakly but can act as an effective short term safe haven against extreme oil price movements. Regarding the currency markets, we put out evidence the gold's capability to act as a strong hedge as well as safe haven against currency market movements. This result has implications for currency investors operating at different time horizons, who want to hedge their exposure to currency swings and with downside risks for those horizons.

**5.3. Intraday Portfolio diversification and hedging ratios**

The aim of our study is to determine the role of gold as a hedge and safe haven asset for the majority of market participants; traders, hedges funds and policy markers. In this context, we evaluate the attractiveness of gold in terms of risk management by taking into account normal (q=2) and turmoil (q=4) movements coming from currencies and oil in different time scales s. Following Kroner and Ng (1998), the optimal weight of gold in a one dollar portfolio of gold/ (currencies or oil) is given by:

$$w_g(s) = \frac{F_{c/o}^q(s) - F_{g,c/o}^q(s)}{F_g^q(s) - 2F_{g,c/o}^q(s) + F_{c/o}^q(s)},$$

(23)

and

$$w_g(s) = \begin{cases} 0 & \text{if} & w_g(s) < 0 \\ w_g(s) & \text{if} & 0 \leq w_g(s) \leq 1 \\ 1 & \text{if} & w_g(s) > 1 \end{cases} \quad (24)$$

where g, o and c denote gold, oil and currency, respectively. $F_{c/o}^q(s)$, $F_g^q(s)$ and $F_{g,c/o}^q(s)$ refer to the qth order detrended fluctuation function of currencies, oil and gold and qth order detrended cross-correlation function between different variables and gold for each time scale, respectively. All these series are estimated through the q-DMCA coefficient. We note that the weight of the currencies (or oil) in the one dollar gold/currencies (or oil) portfolio for different time scales is equal to (1-wg). In addition to the optimal portfolio allocation, investors and market participants seek to minimize the risk of the hedged portfolio and to reduce the cost risk. According to a hedging strategy consisting of holding a long spot position in one unit of currency or oil futures market hedged by a short position of β(s) in the gold futures market (see e.g., Kroner and Sultan, 1993 and Hull, 2011) given by:

$$\beta(s) = \frac{F_{g,c/o}^q(s)}{F_g^q(s)}. \quad (25)$$

**Table 6.** Gold-oil and gold-currency cross-correlations for different time scales and medium fluctuations (q=2)

| s | Gold/Oil | | Gold/EUR | | Gold/GBP | | Gold/JPY | | Gold/CHF | | Gold/CAD | | Gold/AUD | |
|---|---|---|---|---|---|---|---|---|---|---|---|---|---|---|
| | statistic | p-value | statistic | p-value | statistic | p-value | statistic | p-value | statistic | p-value | statistic | p-value | statistic | p-value |
| 20 | 0.0421*** | 0.000 | -0.4165*** | 0.000 | -0.2887*** | 0.000 | -0.4876*** | 0.000 | -0.4731*** | 0.000 | -0.3175*** | 0.000 | -0.3774*** | 0.000 |
| 50 | 0.0391*** | 0.01 | -0.4584*** | 0.000 | -0.2982*** | 0.000 | -0.5160*** | 0.000 | -0.4965*** | 0.000 | -0.3195*** | 0.000 | -0.3706*** | 0.000 |
| 100 | 0.0284 | 0.16 | -0.4648*** | 0.000 | -0.2810*** | 0.000 | -0.5271*** | 0.000 | -0.4933*** | 0.000 | -0.3297*** | 0.000 | -0.3853*** | 0.000 |
| 200 | -0.0149 | 0.67 | -0.5002*** | 0.000 | -0.2999*** | 0.000 | -0.5044*** | 0.000 | -0.5080*** | 0.000 | -0.3517*** | 0.000 | -0.4280*** | 0.000 |
| 398 | 0.0011 | 0.84 | -0.5149*** | 0.000 | -0.3570*** | 0.000 | -0.5342*** | 0.000 | -0.5121*** | 0.000 | -0.4037*** | 0.000 | -0.4787*** | 0.000 |
| 631 | 0.0748* | 0.1 | -0.5178*** | 0.000 | -0.4021*** | 0.000 | -0.5413*** | 0.000 | -0.5081*** | 0.000 | -0.3990*** | 0.000 | -0.4686*** | 0.000 |
| 1000 | 0.0925 | 0.18 | -0.5455*** | 0.000 | -0.4268*** | 0.000 | -0.5577*** | 0.000 | -0.5443*** | 0.000 | -0.3855*** | 0.000 | -0.4832*** | 0.000 |
| 1585 | 0.0421 | 0.69 | -0.5484*** | 0.000 | -0.3943*** | 0.000 | -0.5934*** | 0.000 | -0.5464*** | 0.000 | -0.3933*** | 0.000 | -0.4989*** | 0.000 |
| 1995 | -0.0124 | 0.39 | -0.5313*** | 0.000 | -0.3500*** | 0.000 | -0.5945*** | 0.000 | -0.5204*** | 0.000 | -0.4096*** | 0.000 | -0.4620*** | 0.000 |
| 3162 | -0.0303 | 0.91 | -0.5193*** | 0.000 | -0.2595** | 0.02 | -0.6073*** | 0.000 | -0.4110*** | 0.000 | -0.4033*** | 0.000 | -0.4148*** | 0.000 |

Note: ***, **, and * indicate significance of the coefficients at the 1% level, 5% level, and at 10%, respectively.

**Table 7.** Gold-oil and gold-currency cross-correlations for different time scales and high fluctuations (q=4)

| s | Gold/Oil | | Gold/EUR | | Gold/GBP | | Gold/JPY | | Gold/CHF | | Gold/CAD | | Gold/AUD | |
|---|---|---|---|---|---|---|---|---|---|---|---|---|---|---|
| | statistic | p-value | statistic | p-value | statistic | p-value | statistic | p-value | statistic | p-value | statistic | p-value | statistic | p-value |
| 20 | -0.1477*** | 0.000 | -0.4644*** | 0.000 | -0.2264*** | 0.000 | -0.4001*** | 0.000 | -0.5203*** | 0.000 | -0.2953*** | 0.000 | -0.3739*** | 0.000 |
| 50 | -0.1026*** | 0.000 | -0.4617*** | 0.000 | -0.2582*** | 0.000 | -0.4247*** | 0.000 | -0.5161*** | 0.000 | -0.3209*** | 0.000 | -0.3425*** | 0.000 |
| 100 | -0.1281*** | 0.000 | -0.4396*** | 0.000 | -0.2534*** | 0.000 | -0.3883*** | 0.000 | -0.4700*** | 0.000 | -0.3148*** | 0.000 | -0.3617*** | 0.000 |
| 200 | -0.1801 *** | 0.000 | -0.3862*** | 0.000 | -0.1756*** | 0.000 | -0.3466*** | 0.000 | -0.3771*** | 0.000 | -0.2590*** | 0.000 | -0.3441*** | 0.000 |
| 398 | -0.1488*** | 0.000 | -0.3534*** | 0.000 | -0.2015*** | 0.000 | -0.3689*** | 0.000 | -0.3480*** | 0.000 | -0.2692*** | 0.000 | -0.3117*** | 0.000 |
| 631 | -0.0554* | 0.09 | -0.3977*** | 0.000 | -0.2983*** | 0.000 | -0.3748*** | 0.000 | -0.3716*** | 0.000 | -0.2559*** | 0.000 | -0.2996*** | 0.000 |
| 1000 | -0.0165 | 0.59 | -0.4227*** | 0.000 | -0.2707*** | 0.000 | -0.3559*** | 0.000 | -0.3802*** | 0.000 | -0.2502*** | 0.000 | -0.3255*** | 0.000 |
| 1585 | -0.0486 | 0.35 | -0.3702*** | 0.000 | -0.2516*** | 0.000 | -0.4223*** | 0.000 | -0.3194*** | 0.000 | -0.2634*** | 0.000 | -0.3458*** | 0.000 |
| 1995 | -0.0929* | 0.08 | -0.3153*** | 0.000 | -0.1691** | 0.04 | -0.4030*** | 0.000 | -0.2809*** | 0.000 | -0.2361*** | 0.000 | -0.2711*** | 0.000 |
| 3162 | -0.0339 | 0.72 | -0.2984*** | 0.000 | -0.1167 | 0.8 | -0.4309*** | 0.000 | -0.2101** | 0.03 | -0.3086*** | 0.000 | -0.2752** | 0.03 |

Note: ***, **, and * indicate significance of the coefficients at the 1% level, 5% level, and at 10%, respectively.

**Table 8.** Optimal weights and hedge ratios for different time scales and for q=2

| | s | 20 | 50 | 100 | 200 | 398 | 631 | 1000 | 1585 | 1995 | 3162 |
|---|---|---|---|---|---|---|---|---|---|---|---|
| **Oil** | $w_g$ | 0.9089 | 0.9080 | 0.9067 | 0.8891 | 0.8812 | 0.8956 | 0.9136 | 0.9118 | 0.9067 | 0.9042 |
| | $\beta$ | 0.1250 | 0.1162 | 0.0852 | -0.043 | 0.0029 | 0.1989 | 0.2626 | 0.1274 | -0.0395 | -0.0971 |
| **EUR** | $w_g$ | 0.2464 | 0.2209 | 0.2292 | 0.2241 | 0.1999 | 0.1827 | 0.1708 | 0.1395 | 0.1507 | 0.1297 |
| | $\beta$ | -0.3036 | -0.3307 | -0.3404 | -0.3721 | -0.3754 | -0.3701 | -0.3925 | -0.3802 | -0.3682 | -0.3457 |
| **GBP** | $w_g$ | 0.4021 | 0.4000 | 0.4159 | 0.4104 | 0.3777 | 0.3332 | 0.3017 | 0.3322 | 0.3679 | 0.4049 |
| | $\beta$ | -0.2508 | -0.2588 | -0.2487 | -0.2642 | -0.3043 | -0.3276 | -0.3371 | -0.3199 | -0.2938 | -0.2251 |
| **JPY** | $w_g$ | 0.1562 | 0.1389 | 0.1449 | 0.1477 | 0.1148 | 0.1030 | 0.0976 | 0.0696 | 0.0690 | 0.0701 |
| | $\beta$ | -0.3280 | -0.3470 | -0.3611 | -0.3401 | -0.3531 | -0.3544 | -0.3686 | -0.3929 | -0.3937 | -0.4090 |
| **CHF** | $w_g$ | 0.1516 | 0.1347 | 0.1521 | 0.1624 | 0.1524 | 0.1385 | 0.0974 | 0.0861 | 0.1167 | 0.2209 |
| | $\beta$ | -0.3120 | -0.3259 | -0.3373 | -0.3507 | -0.3499 | -0.3391 | -0.3544 | -0.3506 | -0.3402 | -0.2876 |
| **CAD** | $w_g$ | 0.2962 | 0.2942 | 0.3027 | 0.2878 | 0.2475 | 0.2443 | 0.2695 | 0.2775 | 0.2891 | 0.2912 |
| | $\beta$ | -0.2372 | -0.2382 | -0.2501 | -0.2634 | -0.2926 | -0.2871 | -0.2856 | -0.2960 | -0.3157 | -0.3108 |
| **AUD** | $w_g$ | 0.3608 | 0.3580 | 0.3649 | 0.3345 | 0.2698 | 0.2843 | 0.3397 | 0.3800 | 0.4039 | 0.3848 |
| | $\beta$ | -0.3163 | -0.3088 | -0.3254 | -0.3523 | -0.3721 | -0.3688 | -0.4078 | -0.4417 | -0.4163 | -0.3619 |

**Table 9.** Optimal weights and hedge ratios for different time scales and for q=4

| | s | 20 | 50 | 100 | 200 | 398 | 631 | 1000 | 1585 | 1995 | 3162 |
|---|---|---|---|---|---|---|---|---|---|---|---|
| **Oil** | $w_g$ | 0.9831 | 0.9846 | 0.9833 | 0.9697 | 0.9646 | 0.9726 | 0.9852 | 0.9881 | 0.9861 | 0.9902 |
| | $\beta$ | -1.908 | -1.2158 | -1.5661 | -1.6421 | -1.1207 | -0.3879 | -0.1442 | -0.5513 | -1.1391 | -0.4041 |
| **EUR** | $w_g$ | 0.1512 | 0.1035 | 0.1596 | 0.1370 | 0.1051 | 0.0655 | 0.0663 | 0.0677 | 0.0872 | 0.0649 |
| | $\beta$ | -0.3038 | -0.2773 | -0.2852 | -0.2290 | -0.1882 | -0.2020 | -0.2229 | -0.1818 | -0.1523 | -0.1303 |
| **GBP** | $w_g$ | 0.6049 | 0.5466 | 0.6041 | 0.5509 | 0.4128 | 0.2956 | 0.2424 | 0.2597 | 0.3198 | 0.3579 |
| | $\beta$ | -0.2668 | -0.2767 | -0.2966 | -0.1910 | -0.1750 | -0.2208 | -0.1801 | -0.1710 | -0.1238 | -0.0902 |
| **JPY** | $w_g$ | 0.0215 | 0.0068 | 0.0400 | 0.0329 | 0.0173 | 0.0132 | 0.0367 | 0.0130 | 0.0269 | 0.0137 |
| | $\beta$ | -0.1765 | -0.1858 | -0.1794 | -0.1443 | -0.1497 | -0.1511 | -0.1533 | -0.1884 | -0.1825 | -0.1962 |
| **CHF** | $w_g$ | 0.1665 | 0.1469 | 0.1492 | 0.1287 | 0.1208 | 0.1200 | 0.1408 | 0.1648 | 0.2027 | 0.1655 |
| | $\beta$ | -0.1714 | -0.1825 | -0.1789 | -0.1322 | -0.1358 | -0.1269 | -0.1308 | -0.1481 | -0.1416 | -0.1808 |
| **CAD** | $w_g$ | 0.2628 | 0.2730 | 0.2615 | 0.1729 | 0.1298 | 0.1515 | 0.2288 | 0.3366 | 0.3555 | 0.2907 |
| | $\beta$ | -0.2727 | -0.2497 | -0.2616 | -0.2109 | -0.1686 | -0.1687 | -0.2185 | -0.2776 | -0.2185 | -0.1999 |
| **AUD** | $w_g$ | 0.2016 | 0.2477 | 0.2340 | 0.2192 | 0.2635 | 0.2193 | 0.1524 | 0.1233 | 0.1484 | 0.2654 |
| | $\beta$ | -0.0929 | -0.1290 | -0.1090 | -0.1527 | -0.1809 | -0.2120 | -0.2005 | -0.1590 | -0.2112 | -0.2962 |

Tables 8 and 9 display portfolio weights and hedge ratios for calm and turmoil periods, respectively. Beginning with the portfolio weights, in a 100 USD portfolio of gold and oil futures, the optimal portfolio weight of gold futures for calm period varies from 0.8812 USD (s=398) to 0.9136 USD (s=1000). We deduce that the weight of gold futures in gold and oil futures portfolio remained important and stable as the time scale increased. For gold futures and currency portfolios, weights vary substantially across exchange rates. They range between 12.97% (s=3162) and 24.64% (s=20) for EUR/USD; 30.17% (s=1000) and 41.59% (s=100) for GBP/USD; 6.9% (s=1995) and 15.62% (s=20) for JPY/USD; 8.61% (s=1585) and 22.09% (s=3162) for CHF/USD; 24.43% (s=631) and 30.27% (s=100) for CAD/USD; and 26.98% (s=398) and 40.39% (s=1995) for AUD/USD. These results suggest that i) the weight of gold futures is important in a gold–exchange rate portfolio, especially for a short horizon and ii) it decreased as the time scale increased for the EUR, JPY, and CHF.

The hedge ratio with regard to oil futures fall in the range of -0.0395 (s=1995) to 0.2626 (s=1000). This result suggests that to minimise risk for short hedgers in a 4-week trade, a long position of 1 USD in the oil market should be hedged by a short position of 0.0395 USD in the gold market. However, the hedge ratios for currencies are important; they vary from -0.3036 (s=20) to -0.3925 (s=1000) for EUR/USD; -0.2251 (s=3162) to -0.3371 (s=1000) for GBP/USD; -0.3280 (s=20) to -0.4090 (s=3162) for JPY/USD; -0.2876 (s=3162) to -0.3544 (s=1000) for CHF/USD; -0.2372 (s=20) to -0.3157 (s=1995) for CAD/USD; and -0.3088 (s=50) to -0.4417 (s=1585) for AUD/USD. We deduce that i) the hedge ratio increased when the time scale increased and ii) investors require more gold assets for intraday investments (s=20) to minimise portfolio risk.

In the same way, the optimal portfolio weight and hedge ratio for turmoil periods are presented in Table 9. The empirical results suggest that the weight of gold futures is also important in gold/oil and gold/currency portfolios, except for the JPY. From the results of the hedge ratio, we find that, contrary to calm periods, in turmoil periods, the hedge ratio decreased when the time scale increased for the oil, EUR, GBP, and CAD. This finding implies that to minimise oil and exchange rate risk, especially in less than 1-week trade (for s<631), investors should hold more gold assets in turmoil periods in the case of oil and these three currencies.

On the whole, our results on the usefulness of gold in hedging and safe haven at different investment horizons favor the benefits of including gold futures in a oil futures and currency portfolios for risk management purposes, even though the size of those benefits varies through investment horizons according to specific kinds of portfolios, namely those whose portfolio diversification and hedging ratios were optimally determined.

## 6. Conclusion

In this paper, we proposed a new measure based on the detrending cross-correlation moving average analysis, called the q-DMCA coefficient. This coefficient made more flexible, which depends on the exponent q and the temporal scale s. This feature gives us the opportunity to study the structure of the cross-correlations among several fluctuations. In this work, we limit on average movements (q=2) and extreme market movements (q=4) in order to focus on the capacity of gold to hedge and safe haven against USD depreciation and oil price movements. The feature that the gold price in USD, oil price and the USD value tend to move in opposite directions noted by investors and financial media is checked by a bootstrap test based on surrogated data procedure.

Our results provide evidence of negative and significant average and tail dependence for all time scales between gold and USD exchange rates that is consistent with the gold's role as an effective hedge and safe haven asset. Furthermore, evidence of average independence for all time scales and negative and significant tail dependence between gold and oil for short time scales indicates that gold can be used by investors as a weak hedge and can be used as an effective short term safe-haven asset under exceptional market circumstances.

Extending our analysis to the optimal hedging strategies between gold, currency and oil markets, evidence pointed that in order to reduce the risk for different investment horizons investors should add gold in their portfolios without lowering the anticipated returns of their portfolios.